%% file: nucl-dvcs.tex
\RequirePackage{lineno}                                                %LINE_NUMBERING
\documentclass[prl,nofootinbib,floatfix,twocolumn,showpacs,superscriptaddress]{revtex4}
\usepackage[dvips]{pstcol}
\usepackage{graphicx,amsmath}% Include figure files
\usepackage{dcolumn}% Align table columns on decimal point
\usepackage{bm}% bold math
\usepackage{rotating}
\usepackage{amsmath,amssymb,amsfonts}
\usepackage{color}	

\definecolor{grey}{rgb}{0.75,0.75,0.75}
\definecolor{orange}{rgb}{1.0,0.5,0.5}
\definecolor{brown}{rgb}{0.5,0.25,0.0}
\definecolor{pink}{rgb}{1.0,0.5,0.5}
\definecolor{green}{rgb}{0.,0.5,0.}

\begin{document}

%---------------------- specific definitions------------------
\def\desy{{\sc Desy~}}
\newcommand{\qq}{\ensuremath{Q^2}}

\def\de{\mathrm{d}}
\def\rmT{\mathrm{T}}
\def\rmBH{\mathrm{BH}}
\def\rmDVCS{\mathrm{DVCS}}
\def\rmI{\mathrm{I}}
\def\rmC{\mathrm{C}}
\def\rmb{\mathrm{b}}
\def\rmA{\mathrm{A}}
\def\rmH{\mathrm{H}}
\def\rmunp{\mathrm{UU}}
\def\rmTP{\mathrm{UT}}
\def\rmLU{\mathrm{LU}}
\def\rmUU{\mathrm{UU}}
\def\rmRe{\mathrm{Re}}
\def\rmIm{\mathrm{Im}}
\def\pgate{{\mathcal P}_1(\phi){\mathcal P}_2(\phi)}

\def\CalH{\mathcal{H}}
\def\CalE{\mathcal{E}}
\def\CalHtil{\widetilde{\mathcal{H}}}
\def\CalEtil{\widetilde{\mathcal{E}}}

\def\to{\rightarrow}
\def\gets{\leftarrow}

%----------------------------------------------------------------

\title{
Nuclear--mass dependence of azimuthal beam--helicity and beam--charge asymmetries in 
deeply virtual Compton scattering }

%\collaboration{HERMES Collaboration}

\input{authors}

\date{\today}

\begin{abstract}
%\internallinenumbers               
                                    %LINE_NUMBERING
The nuclear--mass dependence of azimuthal cross section asymmetries
with respect to charge and 
longitudinal polarization of the lepton beam is studied
for hard exclusive electroproduction of real photons.
The observed beam--charge and beam--helicity asymmetries are attributed to the interference between 
the Bethe--Heitler and 
deeply virtual Compton scattering processes.
For various nuclei, the asymmetries are extracted for both coherent and incoherent--enriched regions, which
involve different (combinations of) generalized parton distributions.
For both regions, the asymmetries are compared to those for a free
proton, and 
no nuclear--mass dependence is found.
\end{abstract}


\maketitle

%\runninglinenumbers                                                    %LINE_NUMBERING
%
%----------------------
%\section{Introduction}
%----------------------
%
Lepton scattering experiments constitute an important source of
information for understanding nucleon structure in the context of QCD.
Until recently, this structure was described by two categories of non--perturbative
objects, form factors and parton distribution functions (PDFs),
which have been measured in elastic and deep--inelastic scattering (DIS)
experiments, respectively.
In the last decade, generalized parton distributions (GPDs)~\cite{Muller,Rady,Ji,Pire}
have been recognized as a key concept for the description
of hard exclusive processes.
GPDs offer a multi--dimensional representation of 
the structure of hadrons at the partonic level, correlating the 
longitudinal momentum fraction carried by the parton with its transverse spatial 
coordinate~\cite{Burkardt,Diehl2002,Ralston,Belitsky2002,Burkardt2003}. 
For recent theoretical reviews, see Refs.~\cite{Goeke,Bel02a,Diehl,Belitsky}.

Generalized parton distributions depend 
on the squared four--momentum transfer $t$ to the nucleon and 
on $x$ and $\xi$, which represent respectively the average and half the difference of the
longitudinal momentum fractions carried by the probed parton in initial and final
states. 
Nucleon elastic form factors and parton distribution functions appear as $x$--moments and 
kinematic limits (for $t,\xi\rightarrow 0$) of GPDs, respectively.
The skewness parameter $\xi$ is related to the Bjorken
variable $x_\mathrm{B}=Q^2/(2M\nu)$, 
as $\xi\approx x_\mathrm{B}/(2-x_\mathrm{B})$ in the Bjorken limit
where $Q^2 \rightarrow \infty$ at fixed values of $x_\mathrm{B}$ and $t$.
Here, $M$ is the target mass and $-\qq $ is 
the squared four--momentum of the exchanged virtual photon with energy
$\nu$ in the target rest frame.
Most often discussed are the four \mbox{twist--2} quark--helicity--conserving GPDs for each quark
species in the nucleon: the quark--polarization averaged distributions $H$ and $E$ 
and the quark--polarization related distributions $\widetilde{H}$ and $\widetilde{E}$. 

Among all presently practical
hard exclusive probes, deeply virtual Compton scattering (DVCS),
i.e., the hard exclusive leptoproduction of a real photon,
appears to have the most reliable interpretation in terms of GPDs. 
The final state of the DVCS process
in which the real photon is radiated by a quark 
is intrinsically indistinguishable
from that of the Bethe--Heitler (BH) process
in which a real photon is radiated by the incoming or outgoing lepton.
Access to the DVCS amplitude is provided by interference between the
Bethe--Heitler and DVCS processes,
e.g., via the measurement of the
cross--section asymmetries with respect to the lepton beam helicity and
charge.

This paper reports the first
experimental study of DVCS on nuclear targets.
Nuclei provide a laboratory where, compared to the free nucleon, 
additional information can be obtained on 
GPDs by observing how they become modified in the nuclear environment.
Therefore, studies of nuclear GPDs offer a new opportunity 
to investigate the nature of the nuclear environment.

In lepton--nucleus scattering, two processes can be distinguished for
both DVCS and BH:
(a) the coherent process 
where the electron scatters off the whole nucleus,
which stays intact;
(b) the incoherent process where the  electron scatters
quasi--elastically from an individual nucleon, breaking up the nucleus.

For coherent scattering, 
various DVCS observables have been estimated theoretically
\mbox{~\cite{kirchner-muller:2003,guzey-strikman:2003}}.
In these estimates, 
nuclear GPDs are expressed in terms of 
nucleon GPDs convoluted with the distribution of nucleons in the nucleus.
The $t$ dependence is modeled using nuclear 
elastic form factors.
These models predict an enhancement of the 
beam-charge and beam-helicity asymmetries
for spin--0 and spin--1/2 nuclei compared 
to the case of a free proton.

Recently, coherent DVCS on nuclei has been 
suggested to provide new insights into
the origin of the EMC effect~\cite{Frankfurt:1988,EMC-effect-review,EMC-effect-review-b},
as models that attempt to explain the EMC effect in the forward case 
($t,\xi \rightarrow 0$) also predict
nuclear GPDs that differ from those of a free nucleon (`generalized' EMC effect).
GPD models embodying PDFs that describe the EMC effect observed in inclusive DIS
predict a much larger generalized EMC effect for DVCS observables
~\cite{pire-2001,pire-2004,scopetta-2004,liuti-taneja-2005,guzey-siddikov:2005}. 
In Ref.~\cite{liuti-taneja-2005}, this enhancement is attributed to the transverse
motion of quarks in nuclear targets,
while Ref.~\cite{guzey-siddikov:2005} relates the enhancement to mesonic degrees of freedom 
in hard reactions on nuclei, 
which have been invoked in the `pion excess' models to explain the  EMC effect
in inclusive DIS~\cite{Frankfurt:1988,Arneodo:1992,EMC-effect-review}.
An observable found to be sensitive to  mesonic degrees of freedom is the real part of the DVCS amplitude,
which is predicted to strongly 
depend on the nuclear mass number $A$~\cite{guzey-siddikov:2005}.

Incoherent scattering
is approximated by scattering on free nucleons.
In the kinematic conditions of this experiment, scattering on the
proton dominates due to the fact that the BH process dominates the
single photon production rate and the BH process on the neutron
is suppressed because of the small electromagnetic form factors
compared to those of the proton.
Therefore the asymmetries for nuclei in the incoherent
case are anticipated to be similar to those for the proton.
The role of the neutron contribution was studied in Ref.~\cite{guzey-neutron:2008}.
It was shown to decrease the asymmetries measured in incoherent
nuclear DVCS at larger values of $-t$.
%
%-----------------------
%\subsection{Formalism}
%-----------------------
%
\\\newline\indent 
The cross section for hard exclusive leptoproduction of real photons reads
\begin{eqnarray} 
\frac{\de \sigma}{\de x_\mathrm{B} \de Q^2 \,\de |t| \,\de \phi} &=& \frac{x_\mathrm{B} e^6}{32(2\pi)^4Q^4}\frac{|\rmT|^2}{\sqrt{1+\epsilon^2}} ,
\end{eqnarray}
where $e$ represents the elementary charge, $\epsilon \equiv 2x_\mathrm{B}M/Q$ and
$\rmT$ is the total reaction amplitude.
The azimuthal angle $\phi$ is defined as the angle between the lepton scattering plane
and the photon production plane spanned by the trajectories of the virtual and real photons, following
Ref.~\cite{trento-convention}.
The scattering amplitudes of the DVCS and BH processes add coherently.
The  cross section is then proportional to the 
squared photon--production amplitude written as 

\begin{eqnarray} 
\left|\rmT\right|^2 &=&  \left|\rmT_{\rmDVCS}\right|^2 +  \left|\rmT_{\rmBH}\right|^2  + \rmI ,
\label{eq:eq1}
\end{eqnarray}
where the interference term $\rmI$ is given by
\begin{eqnarray} 
\rmI &=& \rmT_{\rmDVCS}\rmT_{\rmBH}^* +\rmT_{\rmDVCS}^*\rmT_{\rmBH}.
\label{eq:eq1b}
\end{eqnarray}
The BH amplitude $\rmT_\rmBH$ is 
calculable from measured elastic form factors of the (nucleon) nucleus
when modelling the observables for the (in)coherent process.
At leading order in the fine structure constant $\alpha$ 
and for an unpolarized target,
the squared BH amplitude $\left|\rmT_{\rmBH}\right|^2$ is independent of beam polarization 
and lepton charge. 
In contrast, the squared DVCS amplitude $\left|\rmT_{\rmDVCS}\right|^2$ and the interference 
term $\rmI$ depend on the beam helicity, while
the interference term depends also on the lepton charge. 
For a longitudinally polarized lepton beam and unpolarized target, these dependences 
read~\cite{Bel02a}
\begin{eqnarray} 
\left|\rmT_{\rmBH}\right|^2 &=& \frac{K_\rmBH}{\pgate} \, \sum_{n=0}^{2} \, \left[ c_n^\rmBH\cos(n\phi)\right] , \label{eq:BH2}\\
\left|\rmT_{\rmDVCS}\right|^2 &=& \frac{1}{Q^2} \, \Bigl( \, \sum_{n=0}^{2} \left[c_n^\rmDVCS\cos(n\phi)\right] \Bigr.\nonumber \\
&& \hspace{1.2cm} + \, P_\rmb\, s_1^\rmDVCS\sin\phi \, \Bigr) , \label{eq:DVCS}\\
%
%\hspace{1.4cm}
\rmI &=& \frac{-e_\ell K_\rmI }{\pgate} \Bigl( \sum_{n=0}^{3} \left[c_n^\rmI\cos(n\phi)\right]\Bigr.\nonumber\\ 
&& \hspace{1.2cm}  + \, P_\rmb \, \left[ s_1^\rmI\sin \phi + s_2^\rmI\sin (2\phi) \right]\Bigr). 
\label{eq:I}
\end{eqnarray} 
Here, $P_\rmb$ denotes 
the longitudinal beam polarization, $e_\ell$ the beam charge 
in units of the elementary charge, ${\mathcal P}_1(\phi)$ and
${\mathcal P}_2(\phi)$ are the known $\phi$-dependent 
lepton propagators in the BH process,
and the kinematic factors read $K_\rmBH = 1/[x_\mathrm{B}^2t(1+\epsilon^2)^2]$ and 
$K_\rmI = 1/(x_\mathrm{B}yt)$ with $y$ the fraction of the incident lepton energy 
carried by the virtual photon in the target rest frame.
The dependences of the coefficients $c_n$ and $s_n$ on GPDs are
given  in Ref.~\cite{Bel02a}\footnote{Note that the azimuthal angle $\phi$
defined here is different from the one used in Ref.~\cite{Bel02a} 
($\phi=\pi-\phi_{[10]}$).}
for a spin-1/2 target and in Ref.~\cite{Bel2000vk} for a spin-0 target.
For a spin--1/2 target, and within the typical kinematic conditions of this experiment,
the coefficients related to only twist--2 quark GPDs appearing in the interference term 
can be approximated as
\begin{eqnarray}
c_1^\rmI &\propto& F_1 \, \rm{Re} \, \cal{H}, \\
c_0^\rmI &\propto& -\frac{\sqrt{-t}}{Q} \, c_1^\rmI, \label{eqn:c0-c1}\\
s_1^\rmI &\propto& F_1 \, \rm{Im} \, \cal{H},
\end{eqnarray}
where  $\cal{H}$ denotes  the Compton form factor that is  a convolution of the GPD $H$ 
with the hard scattering amplitude,
and $F_1$ is the Dirac form factor.
%
%-------------------------------------------
%\section{Experiment and data selection}
%-------------------------------------------
%
\\\newline\indent
In this paper we present a study of hard exclusive production
of real photons in the reaction 
$e A  \rightarrow e \gamma X$. 
The data were collected with the HERMES spectrometer~\cite{hermes:spectr} 
during the period 1997--2005.
The 27.6 GeV HERA electron or positron beam at DESY was 
scattered off gaseous
hydrogen, helium, nitrogen, neon, krypton, and xenon targets
(see Table~\ref{tab:targets-summary}).
(Results from a deuterium target will be reported
elsewhere~\cite{DC39}.)
The HERA beam was transversely self--polarized due to the Sokolov--Ternov 
mechanism~\cite{sokolov-ternov}.
Longitudinal polarization of the beam was obtained by using a pair of spin
rotators located before and after the interaction region of HERMES.
The beam helicity was reversed every few months.
The beam polarization was measured by two independent HERA 
polarimeters~\cite{polarimeters-a,polarimeters-b}
with a combined fractional systematic uncertainty of up to 3.4\%.
\begin{table}[t]
\begin{center}
\begin{tabular}{l|c|c|c|c}
\hline\hline
$A$ & spin & $L$ (pb$^{-1}$) & $\langle P_b \rangle ^\leftarrow$ &  $\langle P_b \rangle ^\rightarrow$ \\ 
\hline
  H  & 1/2   & 227  & 0.50 & $-$0.51 \\
\hline
He    & 0 & 32 & 0.56 & $-$0.52 \\
\hline
 N     & 1 & 51 & 0.39 & $-$0.40 \\
\hline
 Ne    & 0 & 86 & 0.52 & $-$0.55 \\
\hline
 Kr  & 0  & 77 & 0.43 & $-$0.41 \\
\hline
 Xe   & 0, 1/2, 3/2  & 47 & 0.32 & $-$0.38  \\
\hline
\hline
\end{tabular}
\caption{
Targets used for this analysis, their spins, the corresponding integrated 
luminosity $L$, and the average 
polarization for the two helicity states of the beam.
Note that the xenon target is composed mainly of the isotopes 
$^{129}$Xe (spin--1/2), 
$^{131}$Xe (spin--3/2) and $^{132,134}$Xe (spin--0) with fractional contributions of
26\%, 21\% and 36\%, respectively. 
For all other targets, the admixture of isotopes with spin different from that given in the
table is less than 10\%.
}
\label{tab:targets-summary}
\end{center}
\end{table}
This analysis makes use of the full data set with nuclear targets and a subset of
data with a hydrogen target 
taken in the years 2000 and 2005 corresponding to
approximately 130 pb$^{-1}$ (100 pb$^{-1}$) for the positron (electron) sample.
%
% ttsa paper : 100 (70) for e+ (e-) from 2002-2005 while here: 2000 (e+) and 2005pol+unpol (e-) 
(The results from the full 1996-2005 hydrogen data set
has been reported elsewhere~\cite{DC69}.)
For hydrogen, krypton and xenon targets, data for both positron and electron beams are
available.

A brief description of the event selection is given here.
More details can be found in Refs.~\cite{elli-thesis, bca-paper}.
Events were selected if exactly one photon and one charged track 
identified as the scattered lepton  were detected. 
The hadron contamination in the lepton sample is kept below 1\% 
by combining the information from a transition--radiation detector, a preshower 
scintillator detector, and an electromagnetic calorimeter. 
The kinematic requirements imposed are 1~GeV$^2 < Q^2 <$ 10~GeV$^2$, 0.03 $< x_\mathrm{B} <$ 0.35, 
$\nu <$ 22 GeV, and $W >$ 3 GeV, 
where $W$ is the invariant mass of the virtual--photon nucleon system.
The real photon is identified by a `neutral cluster',
which is defined as 
an energy deposition larger than 5 GeV in the calorimeter and
larger than 1 MeV in the preshower detector, 
and the absence of a corresponding charged track. 
The angular separation $\theta_{\gamma^* \gamma}$ between the virtual and real photons is 
required to be larger than 2 mrad. 
This value is chosen in order to optimize the combined systematic and statistical uncertainties
for the asymmetries
due to the degraded $\phi$  resolution  at low $\theta_{\gamma^* \gamma}$ and the enhanced production 
of real photons on nuclear
targets in the small  $\theta_{\gamma^* \gamma}$ region\footnote{
Note that this value is the only difference from earlier HERMES analyses,
for which $\theta_{\gamma^* \gamma} >$ 5 mrad.}~\cite{Hongxue-thesis}.
An upper bound of 45 mrad is imposed  on this angle in order to improve the signal--to--background 
ratio.

The recoiling system was not detected.
Instead, an `exclusive' sample of events is selected by requiring the squared missing mass 
$M_\mathrm{X}^2= ( q + p - q^\prime)^2$
to correspond within the experimental resolution to the squared proton mass.
Here, $q \, (q^\prime)$ is the four--momentum of the virtual (real) photon 
and $p=(M_\mathrm{p},\overrightarrow{0})$ with $M_\mathrm{p}$ the proton mass.
This selection criterion is chosen by means of a Monte Carlo (MC) simulation
of the missing mass distribution.
%--------------------------- data / MC comparison -----------------------------------------------
%
\begin{figure}[t]
\begin{center}
\includegraphics[width=1.0\columnwidth]{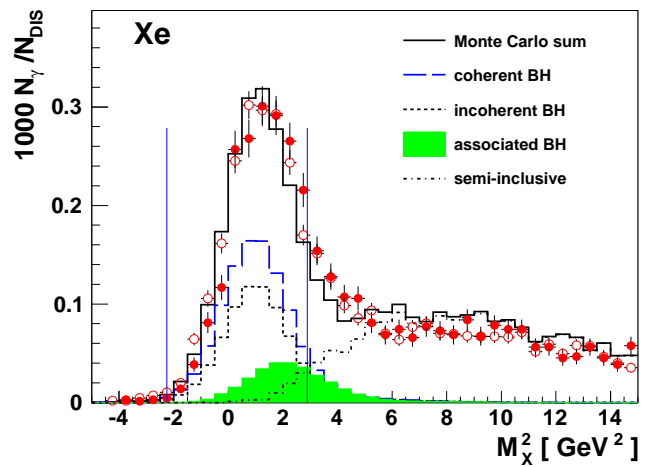}
\end{center}
\caption{(color online) Distributions in squared missing mass from data using positron (filled circles) 
or electron 
(empty circles) beams and a xenon target compared to a MC simulation (solid line).
The latter includes coherent Bethe--Heitler (BH) (dashed line), incoherent BH (short--dashed line) and 
associated BH 
(filled area) processes as well as semi--inclusive background (dash--dotted line). 
The two vertical solid lines enclose the selected exclusive region for the positron data. 
See text for details. 
}
\label{fig:missmass2}
\end{figure}
The result of the simulation is shown in comparison with the experimental data in Fig.~\ref{fig:missmass2}.
In the MC simulation
the expressions in Eqs. 35 and 36 of Ref.~\cite{Bel02a} are used 
for the incoherent BH process.
The simulation also takes into account the incoherent BH process
where a nucleon is excited to a resonant state (known as associated
production) using a parameterization of the total $\gamma^*p$ cross
section for the resonance region from Ref.~\cite{Bra76} and calculating
the individual cross sections for single-meson decay channels, e.g.,
$\Delta^+ \rightarrow p\pi^0$, with the MAID2000 program~\cite{Maid2000}.
For the coherent BH process, the parameterizations of the
form factor for the respective nuclear targets 
are taken from Refs.~\cite{frosch,guzey-neutron:2008}.
%%%%%% 
In addition,
semi--inclusive production of neutral mesons 
(mostly $\pi^0$) is included, where either only one 
photon from the $\pi^0 \rightarrow \gamma\gamma$
decay is detected or these photons
cannot be experimentally resolved.
For this process, the MC generator LEPTO~\cite{Ing97} is used in conjunction with a 
set of JETSET~\cite{Sjo94} fragmentation parameters that had previously been adjusted
to reproduce multiplicity distributions observed by HERMES~\cite{Hil05}. 
Not included in the simulation is 
radiation of more than one photon, which would move events from the peak to the continuum,
nor the DVCS process.
The latter contribution is highly model--dependent. 
In the GPD model used in Ref.~\cite{korotkov02}   
it varies between 10\% and 25\%
of the BH yield for production from a hydrogen target~\cite{thesis-Ye06}.
%
%------------------------------------------------------------------------------------------------- 

The `exclusive region' for the positron data 
is defined as $-(1.5$~GeV)$^2$ $< M_\mathrm{X}^2 < (1.7$~GeV)$^2$, 
where the lower limit is chosen to be 
three times the resolution in $M_\mathrm{X}^2$ from the 
squared proton mass,
and the upper limit 
to optimize the signal--to--background ratio.
Since the $M_\mathrm{X}^2$ spectrum of the electron data is found to be shifted by approximately 0.18~GeV$^2$ 
towards smaller values relative to that of the positron data, 
the exclusive region for electron data is shifted accordingly.
One quarter of the effect of this shift on the results 
presented below is assigned as a contribution to the systematic 
uncertainty.

As the recoiling system was not detected, $t$ is inferred from the measurement of the other 
final--state particles. 
For elastic events, the kinematic relationship between the energy and 
direction of the real photon permits the calculation of $t$ without using the measured energy of the 
real photon, which is the quantity subject to larger uncertainty. 
Thus the value of $t$ is calculated as 
\begin{eqnarray}
t &=& \frac{-Q^2 - 2 \, \nu \, (\nu - \sqrt{\nu^2 + Q^2} \, \cos\theta_{\gamma^* \gamma })}
{1 + \frac{1}{M_\mathrm{p}} \, (\nu - \sqrt{\nu^2 + Q^2} \, \cos\theta_{\gamma^* \gamma })}\label{tc}
\end{eqnarray}
for the exclusive event sample.
The error caused by applying this expression to inelastic events is accounted for in the 
MC simulation that is used to calculate the fractional contribution of background processes
per kinematic bin.
The quantity $-t$ is required to be smaller than 0.7 GeV$^2$.
\begin{figure}[t]
\begin{center}
\includegraphics[width=\columnwidth]{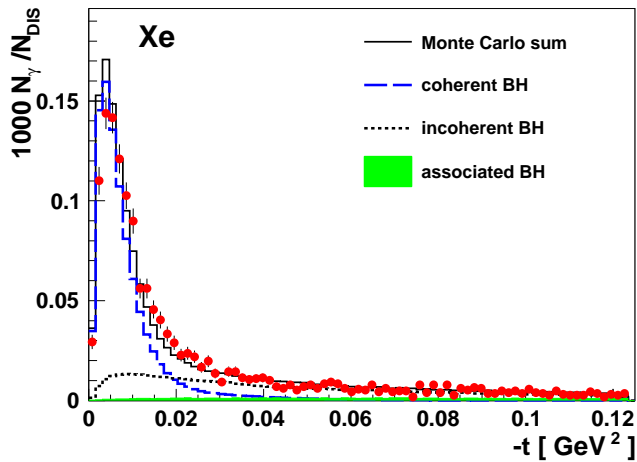}
\end{center}
\caption{
(color online)
Distribution (points) of events selected in the exclusive region
as function of $-t$ compared to a MC simulation (solid line).
The latter includes coherent Bethe--Heitler (BH) (dashed line), 
incoherent BH (dotted line) and associated BH 
(filled area) processes.
Background from semi--inclusive neutral meson production is not included.
}
\label{fig:t} 
\end{figure}
%
%------------------------------------------------------------
%\subsection{separation of coherent / incoherent production}
%------------------------------------------------------------
%

Coherent scattering on nuclear targets is separated from incoherent scattering 
by exploiting its characteristic $t$ dependence.
For both DVCS and BH, coherent scattering occurs at small values of $-t$ and
rapidly diminishes with increasing $|t|$.
However, a complete separation of the two scattering processes is impossible at HERMES.
Coherent--{\it enriched} and incoherent--{\it enriched} samples are selected according to a $-t$
threshold that is chosen to vary with the target such that for each sample 
approximately the same average kinematic conditions 
are obtained for all targets.
The kinematic distributions
of elastic coherent and incoherent processes 
are determined using the MC simulation described above and 
presented in Figs.~\ref{fig:missmass2} and~\ref{fig:t} for xenon,
as an example.
The $t$ distribution of events selected in the exclusive region 
is shown in Fig.~\ref{fig:t} for xenon
together with the simulated contributions of coherent and incoherent processes. 
Tables~\ref{tab:coh} and ~\ref{tab:incoh} summarize the average kinematic conditions
for the various targets for the coherent--enriched and incoherent--enriched samples, respectively,
and give their purities defined as fractions of the total simulated yield.
Also shown for  each sample is the simulated
fractional contribution from the associated BH process.
\begin{table}[t]
\begin{center}
\begin{tabular}{l|c|c|c|c|c|c}
\hline\hline
$A$ &$t$ threshold
&$\langle t \rangle$ &$\,\,\langle x_\mathrm{B} \rangle\,\,$ &$\langle Q^2 \rangle$ & \% of & \% of\\
 & [GeV$^2$] & [GeV$^2$] & & [GeV$^2$] & coh & assoc. \\
\hline
  H    &$-t < 0.033$  & $-0.018$ & 0.070 & 1.81 & -- & 4 \\
\hline
 He    &$-t < 0.036$  & $-0.018$ & 0.072 & 1.83 & 34 & 4 \\
\hline
 N     &$-t < 0.043$  & $-0.018$ & 0.068 & 1.73 & 66 & 3 \\
\hline
 Ne    &$-t < 0.044$  & $-0.018$ & 0.068 & 1.74 & 68 & 3 \\
\hline
 Kr    &$-t < 0.070$  & $-0.018$ & 0.064 & 1.63 & 69 & 3 \\
\hline
 Xe    &$-t < 0.078$  & $-0.018$ & 0.062 & 1.60 & 66 & 4 \\
\hline
\hline
\end{tabular}
\caption {Average kinematics and fractional contributions from coherent 
processes (purity) and associated processes in the coherent--enriched sample 
for the various targets. 
}
\label{tab:coh}
\end{center}
\end{table}
\begin{table}[]
\begin{center}
\begin{tabular}{l|c|c|c|c|c|c}
\hline\hline
$A$ &$t$ threshold
&$\langle t \rangle$ &$\,\,\langle x_\mathrm{B} \rangle\,\,$ &$\langle Q^2 \rangle$ & \% of & \% of\\
 & [GeV$^2$] & [GeV$^2$] & & [GeV$^2$] & incoh & assoc.\\
\hline
  H    &$-t > 0.077$  & $-0.200$ & 0.109 & 2.89 & -- & 20 \\
\hline
 He    &$-t > 0.084$  & $-0.200$ & 0.107 & 2.78 & 61 & 28 \\
\hline
 N     &$-t > 0.083$  & $-0.200$ & 0.113 & 2.93 & 60 & 28 \\
\hline
 Ne    &$-t > 0.075$  & $-0.200$ & 0.111 & 2.92 & 65 & 28 \\
\hline
 Kr    &$-t > 0.067$  & $-0.200$ & 0.108 & 2.84 & 57 & 30 \\
\hline
 Xe    &$-t > 0.060$  & $-0.200$ & 0.107 & 2.86 & 56 & 30 \\
\hline
\hline
\end{tabular}
\caption {Average kinematics and fractional contributions from incoherent 
processes (purity) and associated processes in the incoherent--enriched 
sample for the various targets.
}
\label{tab:incoh}
\end{center}
\end{table}
\begin{table}[]
\begin{center}
\begin{tabular}{l|c|c|c|c|c|c}
\hline\hline
$A$ &$t$ range
&$\langle t \rangle$ &$\,\,\langle x_\mathrm{B} \rangle\,\,$ &$\langle Q^2 \rangle$ & \% of & \% of\\
 & [GeV$^2$] & [GeV$^2$] & & [GeV$^2$] & coh & assoc. \\
\hline
  H    &$0.0 < -t < 0.008$  & $-0.006$ & 0.054 & 1.38 & -- & $< 1$ \\
\hline
 Kr    &$0.0 < -t < 0.010$  & $-0.006$ & 0.053 & 1.37 & 92 & $< 1$ \\
\hline
 Xe    &$0.0 < -t < 0.010$  & $-0.006$ & 0.053 & 1.37 & 92 & $< 1$ \\
\hline
\hline
  H    &$0.008 < -t < 0.020$  & $-0.014$ & 0.069 & 1.75 & -- & 1 \\
\hline
 Kr    &$0.010 < -t < 0.020$  & $-0.014$ & 0.064 & 1.63 & 71 & 2 \\
\hline
 Xe    &$0.010 < -t < 0.020$  & $-0.014$ & 0.062 & 1.67 & 71 & 2 \\
\hline
\hline
\end{tabular}
\caption {Average kinematics and fractional contributions from coherent
processes (purity) 
and associated processes in two $t$ subranges of the coherent--enriched sample 
for hydrogen, krypton and xenon.
}
\label{tab:coh-low-t}
\end{center}
\end{table}
For hydrogen, krypton and xenon, the coherent--enriched region is further explored as a 
function of $t$ (see Table~\ref{tab:coh-low-t}). 

As for these data coherent scattering could not be identified event--by--event, 
kinematic variables that depend on the target mass are always calculated 
using the proton mass.
This does not influence the selection of exclusive events since the values of the
relevant kinematic variables calculated using the proton mass
are highly correlated with those calculated using the actual target mass. 
Also, the calculation of the $t$ value is affected negligibly.
%
%--------------------------------
%\section{Azimuthal asymmetries}
%--------------------------------
%

The full cross section for exclusive production of real photons on unpolarized targets (U) by 
a longitudinally polarized beam (L) 
can be written as
\begin{eqnarray}
\sigma(\phi) &=&\sigma_{\rmUU,0}(\phi)\big[\, 1 + e_\ell {\cal{A}}_{\rmC}(\phi) + P_\rmb
{\cal{A}}_{\rmLU,\rmDVCS}(\phi) \big.\nonumber\\
 && \hspace*{1.6cm} + \, e_\ell P_\rmb \cal{A}_{\rmLU,\rmI}(\phi) \, \big],
\label{eqn:fullxsec}
\end{eqnarray}
where $\sigma_{\rmUU,0}(\phi) = \frac{1}{4}[\sigma^{+\to} + \sigma^{-\to} +
\sigma^{+\gets}+\sigma^{-\gets}]$ 
is the cross section for an unpolarized target averaged over
both beam charges ($+$, $-$) and over both positive ($\rightarrow$)
and negative ($\leftarrow$)  beam helicities.
The  beam--charge asymmetry $\cal{A}_{\rmC}$ and  beam--helicity asymmetries 
$\cal{A}_{\rmLU,\rmDVCS}$ and $\cal{A}_{\rmLU,\rmI}$ are defined in 
Eqs.~\ref{eq:xsection-asym-C}, ~\ref{eq:xsection-asym-LU-DVCS}, and ~\ref{eq:xsection-asym-LU-I},
respectively. 
Each definition is complemented by the corresponding relation to the coefficients
given in Eqs.~\ref{eq:BH2}--\ref{eq:I}:
\begin{eqnarray}
\cal{A}_{\rmC}(\phi) &\equiv& \frac{\sigma^{+\to} - \sigma^{-\to} + \sigma^{+\gets} -
\sigma^{-\gets}} {\sigma^{+\to} + \sigma^{-\to} + \sigma^{+\gets} +
\sigma^{-\gets}} \label{eq:xsection-asym-C} \\
 &=&\frac{-1}{\sigma_{\rmUU,0}(\phi)}
\frac{K_\rmI}{{\cal P}_1(\phi) {\cal P}_2(\phi)} \sum_{n = 0}^{3} c_n^{\rmI}~\cos(n\phi), 
\label{eq:fourier:Ac}
\end{eqnarray}
\begin{eqnarray}
\cal{A}_{\rmLU,\rmDVCS}(\phi) &\equiv& \frac{\sigma^{+\to} + \sigma^{-\to} - \sigma^{+\gets} -
\sigma^{-\gets}} {\sigma^{+\to} + \sigma^{-\to} + \sigma^{+\gets} +
\sigma^{-\gets}} \label{eq:xsection-asym-LU-DVCS}\\
  &=&\frac{1}{\sigma_{\rmUU,0}(\phi)}
\frac{1}{Q^2} s_1^{\rmDVCS}~\sin\phi, 
\label{eq:fourier:Advcs}
\end{eqnarray}
\begin{eqnarray}
\cal{A}_{\rmLU,\rmI}(\phi) &\equiv& \frac{\sigma^{+\to} - \sigma^{-\to} -
\sigma^{+\gets} + \sigma^{-\gets}} {\sigma^{+\to} + \sigma^{-\to} +
\sigma^{+\gets} + \sigma^{-\gets}}   \label{eq:xsection-asym-LU-I}\\
 &=&\frac{-1}{\sigma_{\rmUU,0}(\phi)}
\frac{K_\rmI}{{\cal P}_1(\phi) {\cal P}_2(\phi)} 
\sum_{n=1}^2 s_n^{\rmI} ~\sin(n\phi).
\label{eq:fourier:Aul}
\end{eqnarray}
As the term of Eq.~\ref{eqn:fullxsec} including $\cal{A}_{\rmLU,\rmI}$ 
depends on both beam helicity and beam charge, the DVCS and 
interference beam--helicity asymmetries can be separated.
Such a combined analysis~\cite{ttsa-paper} was performed for hydrogen,
krypton and xenon, where data for both electron and positron
beams are available.
The asymmetries defined in Eqs.~\ref{eq:xsection-asym-C},~\ref{eq:xsection-asym-LU-DVCS}, and
~\ref{eq:xsection-asym-LU-I}
are expanded in terms of the following harmonics in $\phi$:
\begin{eqnarray}
\cal{A}_{\rmC}(\phi) &\simeq& A_{\rmC}^{\cos(0\phi)} 
+ A_{\rmC}^{\cos\phi}\cos\phi  
 \label{eqn:fit-Ac} \\\nonumber
 & + & A_{\rmC}^{\cos(2\phi)}\cos(2\phi) + A_{\rmC}^{\cos(3\phi)}\cos(3\phi) ,\\
\cal{A}_{\rmLU,\rmDVCS}(\phi) &\simeq& A_{\rmLU,\rmDVCS}^{\sin\phi} \sin\phi,
\label{eqn:fit-A_dvcs}\\
\cal{A}_{\rmLU,\rmI}(\phi) &\simeq& A_{\rmLU,\rmI}^{\sin\phi} \sin\phi
+A_{\rmLU,\rmI}^{\sin(2\phi)} \sin(2\phi). \label{eqn:fit-A_I}
\label{eq:amplitudes}
\end{eqnarray}
Using the method of maximum likelihood,
the Fourier coefficients $A$, hereafter called asymmetry amplitudes, are simultaneously extracted
from the event yield that is
proportional to the cross section of Eq.~\ref{eqn:fullxsec}.
Although these asymmetry amplitudes differ somewhat from the coefficients
in Eqs.~\ref{eq:fourier:Ac},~\ref{eq:fourier:Advcs}, and~\ref{eq:fourier:Aul},
they are well defined and can be computed in various GPD models for direct comparison
with data.

For helium, nitrogen and neon, only data with a positron beam were collected.
The single--charge (positron) beam--helicity asymmetry is defined as 
\begin{eqnarray}
\cal{A}_{\rmLU,+}(\phi) &\equiv& \frac{\sigma^{\to} - \sigma^{\gets}}
{\sigma^{\to} + \sigma^{\gets}} \,,  \label{eqn:xsection-asym-LU-single-charge}\\
 &=&  \frac{1}{\sigma_{\rmUU,+}(\phi)}
\frac{1}{Q^2}  s_1^{\rmDVCS} ~\sin\phi \\\nonumber
 &+&\frac{-1}{\sigma_{\rmUU,+}(\phi)}
\frac{e_\ell K_\rmI}{{\cal P}_1(\phi) {\cal P}_2(\phi) }
\sum_{n=1}^2 s_n^\rmI ~\sin(n\phi),
\label{eq:fourier:Aul-single}
\end{eqnarray}
where $\sigma_{\rmUU,+}(\phi) = \frac{1}{2}(\sigma^{\to} + \sigma^{\gets})$. 
In this case the event yield that is proportional to the cross section of
Eq.~\ref{eqn:fullxsec} is fitted by
\begin{eqnarray}
\cal{A}_{\rmLU,+} \simeq& A_{\rmLU,+}^{\sin\phi} \sin\phi +A_{\rmLU,+}^{\sin (2\phi)} \sin (2\phi).
\label{eq:single-charge-amplitude}
\end{eqnarray}
This method does not allow for a separation of squared DVCS amplitude and interference 
term in the beam--helicity asymmetry.
It was used in an earlier
extraction of beam--helicity asymmetries for hydrogen~\cite{hermes:bsa-first}.

In each kinematic bin, the extracted asymmetry amplitudes are corrected for background from 
the decay of semi--inclusively produced neutral mesons, mainly pions. 
The corrected asymmetry amplitude is then obtained as
\begin{eqnarray}
A_\mathrm{corr} &=& \frac{A_\mathrm{raw}-f_\mathrm{semi}\cdot A_\mathrm{semi}}{1-f_\mathrm{semi}},
\label{eq:corr}
\end{eqnarray}
where $A_\mathrm{raw}$ stands for the extracted raw asymmetry amplitude and
 $f_\mathrm{semi}$ and $A_\mathrm{semi}$ for the fractional contribution and corresponding asymmetry amplitude of the 
semi--inclusive background, respectively.
The background contribution $f_\mathrm{semi}$, estimated from MC simulations, 
ranges from $1\%$ to $11\%$ depending on the kinematic conditions
and amounts to 3.5\% on average.
Since the beam--charge--dependent background asymmetry is zero at leading order QED, 
the semi--inclusive background constitutes a dilution for $\cal{A}_{\rmC}$ and effectively also for
$\cal{A}_{\rmLU,\rmI}$, since it cancels in the latter case.
In order to correct $\cal{A}_{\rmLU,\rmDVCS}$ and $\cal{A}_{\rmLU}$ for the 
semi--inclusive background, the size of the corresponding beam--helicity asymmetry is 
extracted from data
by reconstructing neutral pions with a large fractional energy $z=E_{\pi^0}/\nu>0.8$,
as according to MC simulations only these contribute to the exclusive 
region \cite{thesis-Ye06}.
These simulations show that the extracted $\pi^0$ asymmetry does not depend on whether only one or 
both photons are in the acceptance. 
One half of the size of the full background correction is assigned as systematic uncertainty.
Contributions from exclusive $\pi^0$ production were found to be negligible 
at HERMES in a MC simulation
based on a GPD model for exclusive meson production~\cite{VGG} as well as in a data search for 
exclusive $\pi^0$ production~\cite{thesis-Arne}.
Hence this conceivable contribution is not included in the systematic uncertainty.

The asymmetry amplitude $A_\mathrm{corr}$, determined by applying  Eq.~\ref{eq:corr},
is expected to originate from only  elastic and associated production. 
Because essentially nothing is known about the asymmetry for associated production,
no correction is made or uncertainty is assigned for the latter,
but instead associated production is considered to be part of the signal
in this analysis.
The fractional contribution of associated processes is strongly $t$ dependent, 
ranging from 3\% in the lowest $t$ bin to 50\% in the highest $t$ bin,
with little dependence on $A$.
\begin{figure}[t]
\includegraphics[width=1.0\columnwidth]{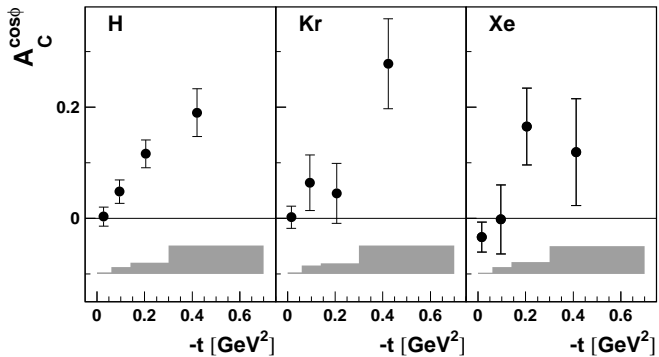}
\caption{
The $\cos\phi$ amplitude of the beam--charge asymmetry for hydrogen, 
krypton and xenon as function of $t$.
The error bars (bands) represent the statistical (systematic) uncertainties.
}
\label{fig:bca-vs-t}
\end{figure}

Effects from detector acceptance, kinematic smearing, finite bin width, 
and from possible detector misalignment 
are estimated using a MC simulation based on the GPD model of Ref.~\cite{Guzey-dual}.
Note that a mistake has been 
found in this GPD model~\cite{Guzey-dual-erratum}; however the model describes well 
the magnitude and kinematic dependences of
previously reported HERMES beam--charge~\cite{ttsa-paper} and preliminary 
beam--helicity asymmetries~\cite{elli-procs-2007}
and thus is considered to be suitable for systematic studies. 
For each bin in $-t$, 
the asymmetry amplitude for hydrogen is 
(i) calculated at the mean kinematic values of a given bin and
(ii) extracted from the reconstructed MC simulation that includes all 
experimental effects.
The difference between these two amplitude values 
is included in the systematic uncertainty.
This uncertainty estimated for hydrogen is applied to all targets.
The validity of this approach was checked using MC simulations based on 
the model of Ref.~\cite{Guzey-dual} that also parameterizes nuclear GPDs.
The systematic uncertainty obtained for the nuclear targets is, within its
statistical uncertainty, in good agreement with that estimated for
hydrogen.

The total systematic uncertainty is dominated by the uncertainty from
the combined contributions of detector acceptance, kinematic smearing, 
finite bin width, and from possible detector misalignment.
This combination is added in quadrature with contributions arising from
the background correction and the relative energy shift between
the $M_\mathrm{X}^2$ spectra of  positron and electron data.
A scale uncertainty of up to 3.4\% arising from 
beam polarimetry is not included
in the systematic uncertainty
for the beam helicity related asymmetries.
Also not included is any contribution due to additional QED vertices, 
as the most significant of these was 
estimated to be negligible in the case of helicity asymmetries~\cite{Afa06}.
\begin{figure}[]
\includegraphics[width=1.0\columnwidth]{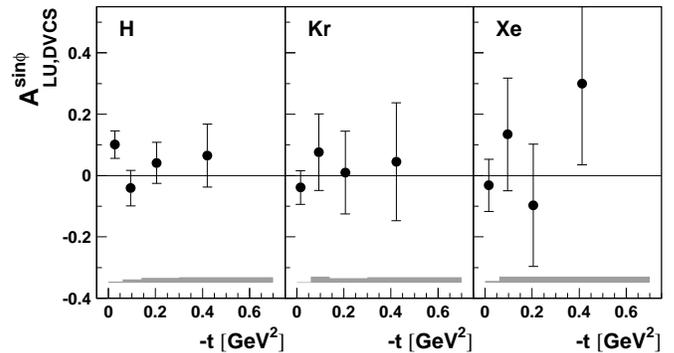}
\caption{
The $\sin\phi$ amplitude of the beam--helicity asymmetry sensitive to the squared DVCS amplitude
for hydrogen, krypton and xenon as function of $t$.
The error bars (bands) represent the statistical (systematic) uncertainties.
This amplitude is subject to an additional 3.4\% maximal scale uncertainty arising from 
beam polarimetry.
}
\label{fig:dvcs-bsa-vs-t}
\end{figure}
\begin{figure}[t]
\includegraphics[width=1.0\columnwidth]{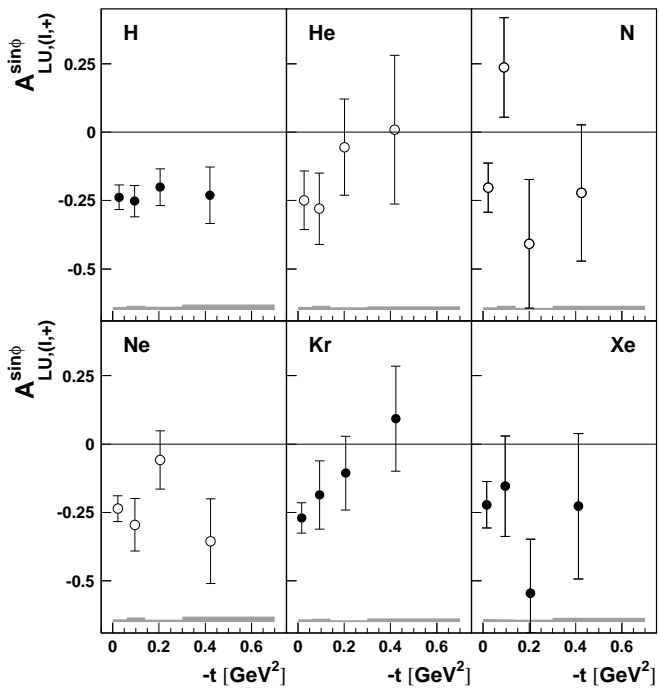}
\caption{
The $t$ dependence of the $\sin\phi$ amplitude of the beam--helicity asymmetry sensitive to the 
interference term, $A_{\rmLU,\rmI}^{\sin\phi}$, for  
hydrogen, krypton and xenon (full symbols) or to a linear combination
of the interference and the squared DVCS amplitude, $A_{\rmLU,+}^{\sin\phi}$, for helium, nitrogen and neon
(open symbols).
The error bars (bands) represent the statistical (systematic) uncertainties.
This amplitude is subject to an additional 3.4\% maximal scale uncertainty arising from 
beam polarimetry.
}
\label{fig:bsa-vs-t}
\end{figure}
%
%
%
%------------------------------------------------------
%\section{Results and comparison to model calculations}
%------------------------------------------------------
%
%
%-------- t-dependence -----------------------------------------------------------------------%
%

In Figs.~\ref{fig:bca-vs-t}--\ref{fig:bsa-vs-t}, the amplitudes of the 
beam--charge and beam--helicity asymmetries,
$A_{\rmC}^{\cos\phi}$, $A_{\rmLU,\rmDVCS}^{\sin\phi}$, $A_{\rmLU,\rmI}^{\sin\phi}$, 
and $A_{\rmLU,+}^{\sin\phi}$,
are shown as functions of $-t$ for unseparated coherent and incoherent production. 
For the nuclear targets,
all other amplitudes in Eqs.~\ref{eqn:fit-Ac}--\ref{eq:amplitudes},~\ref{eq:single-charge-amplitude}
are found to be consistent with zero within 1.5 sigma of the statistical uncertainty.
These other asymmetry amplitudes relate to coefficients
that either embody higher twist quark GPDs or are kinematically suppressed
as, e.g., the amplitude presented in Eq.~\ref{eqn:c0-c1}.

Figure~\ref{fig:bca-vs-t} shows the amplitude $A_{\rmC}^{\cos\phi}$ for hydrogen, krypton, and xenon.
The values for hydrogen 
from this analysis 
are consistent with those extracted previously~\cite{ttsa-paper,bca-paper}.
For hydrogen, krypton and xenon,
the availability of data with both beam charges allows for the separation of the 
azimuthal harmonics of the squared DVCS amplitude and the interference term.
The beam--helicity amplitude $A_{\rmLU,\rmDVCS}^{\sin\phi}$, 
shown in Fig.~\ref{fig:dvcs-bsa-vs-t}, 
is consistent with zero for all three targets over the full $-t$ range.
This is in agreement with the expected suppression of the amplitude.

The beam--helicity amplitudes $A_{\rmLU,\rmI}^{\sin\phi}$ and $A_{\rmLU,+}^{\sin\phi}$
shown in Fig.~\ref{fig:bsa-vs-t}, are substantial  for all targets.
For helium, nitrogen and neon, where only positron beam data are available, this amplitude 
receives also contributions from the squared--DVCS term. 
However, as the latter amplitude is expected to be suppressed
and found to be so for other targets in Fig.~\ref{fig:dvcs-bsa-vs-t},
its contribution is assumed to be small here.
%
%-------- A-dependence -----------------------------------------------------------------------%
%

\begin{figure}[t]
\includegraphics[width=1.0\columnwidth]{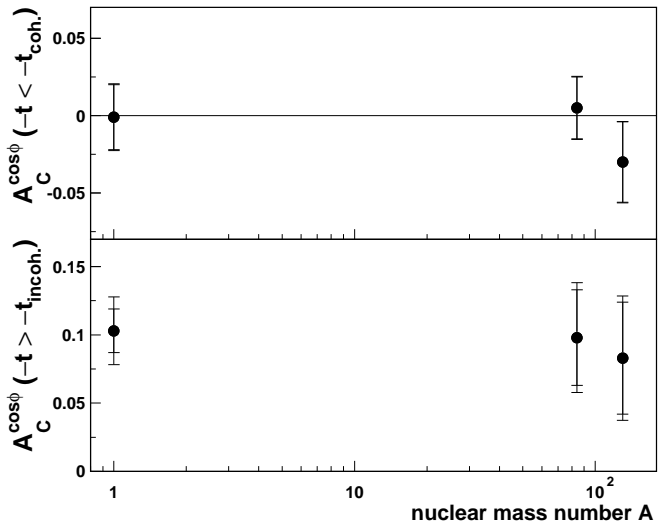}
\caption{
Nuclear--mass dependence of the $\cos\phi$ amplitude of the  beam--charge asymmetry 
for the coherent--enriched (upper panel)
and incoherent--enriched (lower panel) data samples for hydrogen, krypton and xenon.
The coherent--enriched samples have a purity of about 67\% and the incoherent--enriched
samples a purity of about 60\%.
The inner error bars represent the statistical uncertainty and the full bars the quadratic sum of
statistical and systematic uncertainties.
}
\label{fig:bca-vs-A}
\end{figure}

\begin{figure}[t]
\includegraphics[width=1.0\columnwidth]{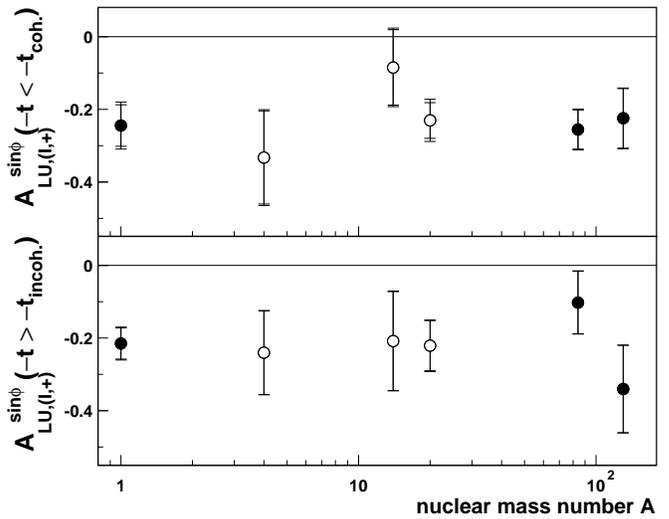}
\caption{
Nuclear--mass dependence of the $\sin\phi$ amplitude of the beam--helicity asymmetry
for the coherent--enriched (upper panel)
and incoherent--enriched (lower panel) data samples.
See Fig.~\ref{fig:bsa-vs-t} for the meaning of open and full circles.
The coherent--enriched samples have a purity of about 67\% except for He with 34\%, and
the incoherent--enriched samples a purity of about 60\%.
The inner error bars represent the statistical uncertainty and the full bars the quadratic sum of
statistical and systematic uncertainties.
This amplitude is subject to an additional 3.4\% maximal scale uncertainty arising from
beam polarimetry.
}
\label{fig:bsa-vs-A}
\end{figure}

The nuclear--mass dependence of the azimuthal beam--charge and beam--helicity 
asymmetries is presented 
separately for the coherent and incoherent--enriched samples
in Figs.~\ref{fig:bca-vs-A} and~\ref{fig:bsa-vs-A}.
The $\cos\phi$ amplitude of the beam--charge asymmetry
is consistent with zero for the coherent--enriched samples for all three targets,
while it is about 0.1 for the incoherent--enriched samples without showing any 
dependence on the target mass within uncertainties. 
The $\sin\phi$ amplitude of the beam--helicity asymmetry shown in Fig.~\ref{fig:bsa-vs-A}
has values of about $-0.2$ for both the coherent and incoherent--enriched samples
without showing any dependence on $A$ within uncertainties.
In order to quantify nuclear effects, the asymmetry amplitudes for nuclear targets
are compared to those for a free proton.
The ratio  
$R_{\rm{LU}} = A_{\rmLU,\rm(I,+),\rmA}^{\sin\phi}/A_{\rmLU,\rmI,\rmH}^{\sin\phi}$
of the nuclear--to--hydrogen beam--helicity asymmetry amplitudes
averaged over all targets, 
is found to be $0.91\pm0.19$ for the coherent--enriched sample and 
$0.93\pm0.23$ for the incoherent--enriched sample, both of which are compatible with unity.

For incoherent scattering, the asymmetry for nuclei is expected to be 
similar to that for hydrogen aside from effects of the nuclear environment,
since scattering on a proton dominates.
Neglecting the neutron contribution, the value of $R_{\rm{LU}}$ for incoherent 
scattering is expected to be unity~\cite{guzey-strikman:2003}.
In Ref.~\cite{guzey-neutron:2008},
the neutron contribution to incoherent nuclear DVCS is taken into account and 
$R_{\rm{LU}}$ is predicted to be 
between 0.66 and 0.74 at $t = -0.2$ GeV$^2$.
Within the experimental uncertainties, 
the measured ratio $R_{\rm{LU}} = 0.93 \pm 0.23$ 
agrees with both the expected suppression of the neutron contribution in 
incoherent scattering on nuclei and with the prediction of Ref.~\cite{guzey-neutron:2008}.

The results for the coherent--enriched samples can be compared to predictions
based on simple models for nuclear GPDs 
that express them in terms of nucleon GPDs 
~\cite{kirchner-muller:2003,guzey-strikman:2003}.
Within this approach, nuclear beam--charge and beam--helicity asymmetries are 
predicted to be essentially independent of $A$ for heavier nuclei.
Compared to the free proton asymmetry,  
nuclear beam--charge and beam--helicity asymmetries are expected 
to be enhanced for spin--0 and spin--1/2 nuclei. 
This predicted enhancement is based on the model--independent observation
that DVCS takes place either  on a proton or neutron in the nuclear target,
while BH occurs predominantly only on a proton. 
The ratio $R_{\rm{LU}}$ of the nuclear--to--hydrogen beam--helicity asymmetry amplitudes
has been estimated in Ref.~\cite{kirchner-muller:2003}
for the pure coherent process 
to be about $5/3$ for spin--0 and spin--1/2 nuclei with Z=N,  
essentially independent of $A$.
This value arises from the ratio of squared charges for an isoscalar to an
isodublet state and the observation that for the valence quark PDFs
$d/u=1/2$ in the kinematics of this experiment.
For spin--1 nuclei,  $R_{\rm{LU}}$ is predicted to be unity.
Ref.~\cite{kirchner-muller:2003} also formulates a GPD model.
Considering only leading twist GPDs and valence quark contributions,
the predicted value of $A_{\rm{LU,I}}^{\sin\phi}$ for hydrogen is $-0.26$ 
for the kinematic condition 
$ t = -0.2$ GeV$^2$, $ Q^2 = 2.5$  GeV$^2$ and
$x_\mathrm{B} = 0.12$. 
Including sea quark contributions, twist--3 corrections and varying the main model
parameters, the predicted amplitude is in the range $0.16 \leq |A_{\rm{LU,I}}^{\sin\phi}| \leq 0.37$.  
In Ref.~\cite{guzey-strikman:2003}, a somewhat more elaborated calculation is presented
where nuclear GPDs are expressed in terms of nucleon GPDs convoluted with the distribution
of nucleons in the nucleus thereby accounting for nuclear binding.
Within this approach, the ratio $R_{\rm{LU}}$ 
is predicted to be about 1.8 for neon and krypton for the kinematic condition 
$ t = -0.018$ GeV$^2$, $ Q^2 = 1.58$  GeV$^2$ and $x_\mathrm{B} = 0.10$. 

\begin{figure}[t]
\includegraphics[width=1.0\columnwidth]{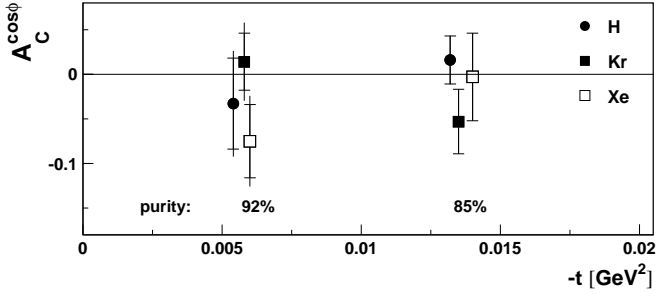}
\caption{
The $\cos\phi$ amplitude of the beam--charge asymmetry for hydrogen, krypton
and xenon as function of $t$.
The inner error bars represent the statistical uncertainty and the full bars the quadratic sum of
statistical and systematic uncertainties.
The purity of the coherent--enriched Kr and Xe samples is indicated for the two $t$ bins.
}
\label{fig:bca-vs-purity}
\end{figure}

\begin{figure}[]
\includegraphics[width=1.0\columnwidth]{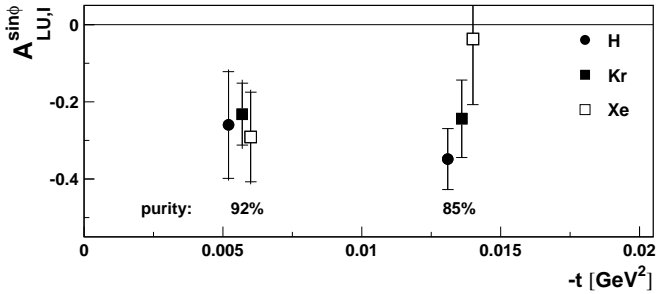}
\caption{
The $\sin\phi$ amplitude of the beam--helicity asymmetry sensitive to the interference term
for hydrogen, krypton
and xenon as function of $t$.
This amplitude is subject to an additional 3.4\% maximal scale uncertainty arising from 
beam polarimetry.
Otherwise as Fig.8.
}
\label{fig:bsa-vs-purity}
\end{figure}

The nuclear beam--helicity amplitudes shown in 
Fig.~\ref{fig:bsa-vs-A} (upper panel) supports the predicted independence of $A$ for heavier targets. 
They do not support the anticipated enhancement of the asymmetries compared to the free
proton asymmetries for  spin--0 and spin--1/2 nuclei.
However, 
the measured amplitude for the coherent--enriched sample receives
contributions from incoherent scattering which 
is expected to diminish $R_{\rm{LU}}$.
The value $R_{\rm{LU}} = 0.91\pm0.19$ for the coherent--enriched samples should therefore be compared
to a prediction involving a mixture of asymmetry amplitudes for coherent and incoherent
processes.
For an average purity of 67\% for the coherent--enriched samples of nitrogen to xenon
(see Table~\ref{tab:coh}) and assuming that the asymmetry from the incoherent portion 
of the yield is the same as for hydrogen,  
the predicted ratio $R_{\rm{LU}}=5/3$ for the pure coherent process becomes 1.45.
In Ref.~\cite{guzey-strikman:2003}, both coherent and incoherent scattering have been 
considered and the predicted ratio $R_{\rm{LU}}$ of about 1.8 for the pure coherent process 
becomes about 1.6.
These values exceed the measured ratio by more than three standard deviations of the 
total experimental uncertainty.
As shown in Figs.~\ref{fig:bca-vs-purity} and ~\ref{fig:bsa-vs-purity},
for coherent--enriched samples
both beam--charge and beam--helicity amplitudes for hydrogen, krypton and xenon are
essentially independent of $t$ within uncertainties.

In Ref.~\cite{guzey-siddikov:2005}, mesonic degrees of freedom were 
also considered in the description of coherent scattering on nuclei
and in the explanation of the generalized EMC effect.
Such a contribution is predicted to significantly enhance the real part of the DVCS 
amplitude, which translates
into a strong nuclear--mass dependence of the beam--charge asymmetry.
In the absence of meson exchange, this asymmetry is expected to be 
essentially independent of $A$ for heavier nuclei.
The nuclear beam--charge amplitudes shown in 
Fig.~\ref{fig:bca-vs-A} (upper panel) do not show any enhancement about
the amplitude for the free proton and do not exhibit any dependence \mbox{on $A$.}
%
%
%---------------------
%\section{summary} 
%---------------------
%
\\\newline\indent
In summary, the nuclear--mass dependence of azimuthal beam--helicity asymmetries in 
electroproduction of real 
photons is measured for the first time for targets ranging from hydrogen to xenon. 
For hydrogen, krypton and xenon, data were taken with both beam charges
and beam helicities allowing a separation of
the $\sin\phi$ amplitude of the squared DVCS and the interference terms.
Also, the $\cos\phi$ amplitude of the beam--charge asymmetry  has been evaluated for those targets.
This amplitude is consistent with earlier 
measurements for hydrogen~\cite{ttsa-paper,bca-paper}.
For the coherent--enriched data sample, the $\cos\phi$ amplitude is found to be consistent 
with zero for all nuclear targets, while it amounts to 0.1 for the incoherent--enriched
data sample, in either case not exhibiting any dependence on $A$ within 
experimental uncertainties.

The $\sin\phi$ amplitude of the beam--helicity asymmetry sensitive to the  
squared DVCS amplitude is consistent with zero for all targets.
The $\sin\phi$ amplitude of the beam--helicity asymmetry sensitive to the  
interference term is significantly non--zero with a value of
about $-0.2$ for both the coherent and incoherent--enriched samples  
without showing any dependence on $A$ within uncertainties.
These amplitudes are compared to those of a free proton. 
The ratio $R_{\rmLU} = A_{\rmLU,\rm(I,+),\rmA}^{\sin\phi}/A_{\rmLU,\rmI,\rmH}^{\sin\phi}$ is found to be 
$0.91\pm0.19$ for the coherent--enriched sample and 
$0.93\pm0.23$ for the incoherent--enriched sample, both of which are compatible with unity.

No nuclear--mass dependence of the
beam--charge and beam--helicity asymmetries is observed for heavier nuclei, 
in agreement with the general feature of 
models that approximate nuclear GPDs by a sum of
nucleon GPDs convoluted with the distribution of nucleons in the nucleus.
The data do not support the enhancement of nuclear asymmetries
compared to the free proton asymmetries
for coherent scattering on spin--0 and spin--1/2 nuclei as anticipated by
various models~\cite{kirchner-muller:2003,guzey-strikman:2003,guzey-siddikov:2005}.
%
%
%
%------------------------ acknowledgments ------------------------------------
%
\\\newline\indent
\begin{acknowledgments} 
We thank M. Diehl, S. Liuti and M. Siddikov for many interesting discussions.
We gratefully acknowledge the \desy\ management for its support and the staff
at \desy\ and the collaborating institutions for their significant effort.
This work was supported by the FWO-Flanders and IWT, Belgium;
the Natural Sciences and Engineering Research Council of Canada;
the National Natural Science Foundation of China;
the Alexander von Humboldt Stiftung;
the German Bundesministerium f\"ur Bildung und Forschung (BMBF);
the Deutsche Forschungsgemeinschaft (DFG);
the Italian Istituto Nazionale di Fisica Nucleare (INFN);
the MEXT, JSPS, and G-COE of Japan;
the Dutch Foundation for Fundamenteel Onderzoek der Materie (FOM);
the U.K.~Engineering and Physical Sciences Research Council, 
the Science and Technology Facilities Council,
and the Scottish Universities Physics Alliance;
the U.S.~Department of Energy (DOE) and the National Science Foundation (NSF);
the Russian Academy of Science and the Russian Federal Agency for 
Science and Innovations;
the Ministry of Economy and the Ministry of Education and Science of 
Armenia;
and the European Community-Research Infrastructure Activity under the
FP6 ``Structuring the European Research Area'' program
(HadronPhysics, contract number RII3-CT-2004-506078).
\end{acknowledgments}

\end{document}

%% file: authors.tex
% Group addresses by affiliation; use superscriptaddress for long
% author lists, or if there are many overlapping affiliations.
% For Phys. Rev. appearance, change preprint to twocolumn.
% Choose pra, prb, prc, prd, pre, prl, prstab, or rmp for journal
%  Add 'draft' option to mark overfull boxes with black boxes
%  Add 'showpacs' option to make PACS codes appear
%  Add 'showkeys' option to make keywords appear

%\documentclass[aps,prl,twocolumn,superscriptaddress]{revtex4}
%\documentclass[aps,prl,preprint,superscriptaddress]{revtex4}
%\documentclass[aps,prl,preprint,groupedaddress]{revtex4}

% Enable use of graphics

%\usepackage{graphicx}

% You should use BibTeX and apsrev.bst for references
% Choosing a journal automatically selects the correct APS
% BibTeX style file (bst file), so only uncomment the line
% below if necessary.

%\bibliographystyle{apsrev}

% ===========================================================================

% List of Institute Addresses 

\def\groupargonne{\affiliation{Physics Division, Argonne National Laboratory, Argonne, Illinois 60439-4843, USA}}
\def\groupbari{\affiliation{Istituto Nazionale di Fisica Nucleare, Sezione di Bari, 70124 Bari, Italy}}
\def\groupbeijing{\affiliation{School of Physics, Peking University, Beijing 100871, China}}
\def\groupcolorado{\affiliation{Nuclear Physics Laboratory, University of Colorado, Boulder, Colorado 80309-0390, USA}}
\def\groupdesy{\affiliation{DESY, 22603 Hamburg, Germany}}
\def\groupzeuthen{\affiliation{DESY, 15738 Zeuthen, Germany}}
\def\groupdubna{\affiliation{Joint Institute for Nuclear Research, 141980 Dubna, Russia}}
\def\grouperlangen{\affiliation{Physikalisches Institut, Universit\"at Erlangen-N\"urnberg, 91058 Erlangen, Germany}}
\def\groupferrara{\affiliation{Istituto Nazionale di Fisica Nucleare, Sezione di Ferrara and Dipartimento di Fisica, Universit\`a di Ferrara, 44100 Ferrara, Italy}}
\def\groupfrascati{\affiliation{Istituto Nazionale di Fisica Nucleare, Laboratori Nazionali di Frascati, 00044 Frascati, Italy}}
\def\groupgent{\affiliation{Department of Subatomic and Radiation Physics, University of Gent, 9000 Gent, Belgium}}
\def\groupgiessen{\affiliation{Physikalisches Institut, Universit\"at Gie{\ss}en, 35392 Gie{\ss}en, Germany}}
\def\groupglasgow{\affiliation{Department of Physics and Astronomy, University of Glasgow, Glasgow G12 8QQ, United Kingdom}}
\def\groupillinois{\affiliation{Department of Physics, University of Illinois, Urbana, Illinois 61801-3080, USA}}
\def\groupmichigan{\affiliation{Randall Laboratory of Physics, University of Michigan, Ann Arbor, Michigan 48109-1040, USA }}
\def\groupmoscow{\affiliation{Lebedev Physical Institute, 117924 Moscow, Russia}}
\def\groupnikhef{\affiliation{National Institute for Subatomic Physics (Nikhef), 1009 DB Amsterdam, The Netherlands}}
\def\groupstpetersburg{\affiliation{Petersburg Nuclear Physics Institute, Gatchina, Leningrad region 188300, Russia}}
\def\groupprotvino{\affiliation{Institute for High Energy Physics, Protvino, Moscow region 142281, Russia}}
\def\groupregensburg{\affiliation{Institut f\"ur Theoretische Physik, Universit\"at Regensburg, 93040 Regensburg, Germany}}
\def\grouprome{\affiliation{Istituto Nazionale di Fisica Nucleare, Sezione Roma 1, Gruppo Sanit\`a and Physics Laboratory, Istituto Superiore di Sanit\`a, 00161 Roma, Italy}}
\def\grouptriumf{\affiliation{TRIUMF, Vancouver, British Columbia V6T 2A3, Canada}}
\def\grouptokyo{\affiliation{Department of Physics, Tokyo Institute of Technology, Tokyo 152, Japan}}
\def\groupamsterdam{\affiliation{Department of Physics, VU University, 1081 HV Amsterdam, The Netherlands}}
\def\groupwarsaw{\affiliation{Andrzej Soltan Institute for Nuclear Studies, 00-689 Warsaw, Poland}}
\def\groupyerevan{\affiliation{Yerevan Physics Institute, 375036 Yerevan, Armenia}}
\def\groupnone{\noaffiliation}

% Set Institute Order 

\groupargonne
\groupbari
\groupbeijing
\groupcolorado
\groupdesy
\groupzeuthen
\groupdubna
\grouperlangen
\groupferrara
\groupfrascati
\groupgent
\groupgiessen
\groupglasgow
\groupillinois
\groupmichigan
\groupmoscow
\groupnikhef
\groupstpetersburg
\groupprotvino
\groupregensburg
\grouprome
\grouptriumf
\grouptokyo
\groupamsterdam
\groupwarsaw
\groupyerevan

% List of Authors 

\author{A.~Airapetian} \groupgiessen \groupmichigan
\author{N.~Akopov}  \groupyerevan
\author{Z.~Akopov}  \groupdesy
\author{M.~Amarian}  \groupzeuthen
\author{E.C.~Aschenauer\footnote{Now at: Brookhaven National Laboratory, Upton, New York 11772-5000, USA}}  \groupzeuthen 
\author{W.~Augustyniak}  \groupwarsaw
\author{R.~Avakian}  \groupyerevan
\author{A.~Avetissian}  \groupyerevan
\author{E.~Avetisyan}  \groupdesy
\author{B.~Ball}  \groupmichigan
% ...

\collaboration{The HERMES Collaboration} \noaffiliation